\documentclass[
    ,final            
  ,numberedheadings 
  ]
  {aipproc}
 \usepackage{amstext}
 \usepackage{amsxtra}
  \newcommand{\bs}{\boldsymbol}

\layoutstyle{6x9}

\begin{document}

\title{The atomic Bose gas in Flatland}


\author{Z.~Hadzibabic}{
  address={Laboratoire Kastler Brossel,
  Ecole normale supérieure, 24 rue Lhomond, 75005 Paris, France}
}
\author{P.~Krüger}{
  address={Laboratoire Kastler Brossel,
  Ecole normale supérieure, 24 rue Lhomond, 75005 Paris, France}
}\author{M.~Cheneau}{
  address={Laboratoire Kastler Brossel,
  Ecole normale supérieure, 24 rue Lhomond, 75005 Paris, France}
}\author{B.~Battelier}{
  address={Laboratoire Kastler Brossel,
  Ecole normale supérieure, 24 rue Lhomond, 75005 Paris, France}
}
\author{J.~Dalibard}{
  address={Laboratoire Kastler Brossel,
  Ecole normale supérieure, 24 rue Lhomond, 75005 Paris, France}
}

\begin{abstract}
We describe a recent experiment performed with rubidium atoms
($^{87}$Rb), aiming at studying the coherence properties of a
two-dimensional gas of bosonic particles at low temperature. We
have observed in particular a Berezinskii--Kosterlitz--Thouless
(BKT) type crossover in the system, using a matter wave
heterodyning technique.  At low temperatures, the gas is
quasi-coherent on the length scale set by the system size. As the
temperature is increased, the loss of long-range coherence
coincides with the onset of the proliferation of free vortices, in
agreement with the microscopic BKT theory.
\end{abstract}

\maketitle


\section{Introduction}

The type of phase transitions that a physical system can undergo
depends crucially on its dimensionality. In a three-dimensional
(3D) system, one usually observes the emergence of a long range
order below a critical temperature. The corresponding order
parameter can be constant in space, like the macroscopic wave
function of a 3D Bose-Einstein condensate (BEC) or the
magnetization in a 3D ferromagnetic material. In a two-dimensional
fluid, long-range order is destroyed by thermal fluctuations at
any finite temperature $T$. In particular, for a 2D uniform ideal
gas of identical bosons it is well known that BEC cannot occur at
a non-zero temperature.

However in an interacting 2D Bose gas, there exists a certain
critical temperature $T_c$ at which a phase transition takes
place. Below $T_c$ correlations of the order parameter decay only
algebraically in space and the system is superfluid. Above $T_c$
the correlations decay exponentially and superfluidity is lost.
The Berezinskii-Kosterlitz-Thouless (BKT) paradigm
\cite{Berezinskii,KT73} associates this transition with the
emergence of a topological order embodied in the pairing of
vortices with opposite circulations. For $T<T_c$ vortices are only
found as part of bound pairs, whereas for $T>T_c$ unbound vortices
proliferate and cause the exponential decay of the correlations of
the order parameter.

This behavior is characteristic of a variety of 2D phenomena
including the superfluidity in liquid helium films
\cite{Bishop78}, the superconducting transition in arrays of
Josephson junctions \cite{Resnick81}, and the collision physics of
2D atomic hydrogen \cite{Safonov98}. Here we report experimental
evidence for BKT-type phenomenon in harmonically trapped dilute
atomic gases. Specifically, we analyze series of interference
patterns between independent quasi 2D gases of $^{87}$Rb atoms. We
identify a temperature region close to the predicted $T_c$, around
which the phase coherence of the gas evolves rapidly with $T$. For
temperatures above this transition region, phase defects indeed
proliferate. Our observations thus pave the way to a direct,
quantitative confirmation of the microscopic basis of the BKT
theory.

The paper is organized as follows. We first present
(section~\ref{sec:production}) our experimental configuration in
order to set the scale of parameters of our quasi-2D gas. We then
review the relevant known results for homogeneous
(section~\ref{sec:homogeneous}) and trapped
(section~\ref{sec:trapped}) Bose gases in two dimensions. In
section~\ref{experiment} we present our main experimental results
concerning the observation of a BKT type crossover in a 2D gas of
rubidium atoms.

\section{Experimental system}
 \label{sec:production}
Our experimental setup has been described in detail in
\cite{Stock05,Hadzibabic06} (see \cite{Other_exps} for other
recent experiments performed with quasi-2D atomic gases). We start
our experiment with a magnetically trapped, quantum degenerate 3D
cloud of $^{87}$Rb atoms. A 1D optical lattice with a period of $d
= 3 \,\mu$m along the vertical direction $z$ is used to split the
3D gas into two independent clouds and to compress them into the
2D regime.

The two clouds form parallel, elongated 2D strips, characterized
by the harmonic trapping frequencies in the $xy$ plane
$\omega_x/(2\pi)=11\,$Hz and  $\omega_y/(2\pi)=130\,$Hz. The
oscillation frequency in the lattice potential along the $z$
direction is $\omega_z/(2\pi)=3.6$~kHz, and the tunneling between
the two planes is negligible on the time-scale of the experiment.

The maximal number of atoms in each planar quasi-condensate is
$\sim 5\times 10^4$. The Thomas-Fermi approximation presented in
section~\ref{sec:trapped} yields $\rho_c=10^{10}$~cm$^2$ for the
central density of each quasi-condensate, and $2R_x=120\,\mu$m and
$2R_y=10\,\mu$m for the $x$ and $y$ lengths of the strips,
respectively. The corresponding chemical potential is $\mu/ h =
1.8$ kHz. Since $\mu <\hbar \omega_z$, the motion along $z$ is
frozen out.

In quasi-2D, the chemical potential can be written as
 \begin{equation}
\mu=\frac{\hbar^2}{M}\;\tilde{g} \rho_c \ ,
 \label{interaction}
 \end{equation}
where $M$ is the atomic mass, and the dimensionless parameter
$\tilde g$ can be expressed in terms of the 3D scattering length
$a=5.2$~nm and the extension of the ground state along the lattice
direction $a_z=\sqrt{\hbar/(M\omega_z)}= 0.17\,\mu$m
 \begin{equation}
 \tilde g \simeq \sqrt{8\pi}\;\frac{a}{a_z} \simeq 0.15\ .
 \label{tildeg}
 \end{equation}
Since $\tilde g<1$, each planar cloud is a weakly interacting
system. The healing length $\xi=\hbar/\sqrt{2M\mu}=1/\sqrt{2\tilde
g\rho_c}\simeq 0.2\,\mu$m gives the characteristic size of the
core of a vortex, and the quantity $\pi\;/\;2\tilde g \simeq 10$
can be interpreted as the number of atoms in the area $\pi \xi^2$
that are missing in the vortex core. Note that strictly speaking
$\tilde g$ also depends on the relevant energy set by the chemical
potential $\mu$ \cite{Adhikari86,Petrov00,Petrov01}, but for our
experimental parameters this dependence may be neglected.

\section{The homogeneous 2D gas}
 \label{sec:homogeneous}
In this section we give a brief summary of the known results
concerning a 2D gas of bosonic particles. Consider first an ideal
gas of $N$ bosons at temperature $T$, confined in a square box of
size $L^2$. Using the Bose-Einstein distribution and assuming a
smooth variation of the population of the various energy states,
we find a relation between the spatial density $\rho=N/L^2$, the
thermal wavelength $\lambda=h/(2\pi M k_B T)^{1/2}$ and the
chemical potential $\mu$:
 \[
\rho \lambda^2 = - \ln \left( 1-e^{\mu/(k_BT)}  \right)\ .
 \]
This relation allows one to derive the value of $\mu$ for any
value of the degeneracy parameter $\rho \lambda^2$. It indicates
that no condensation takes place in 2D, contrary to the 3D case.
In the latter case the relation between $\rho^{(3D)} \lambda^3$
and $\mu$ ceases to admit a solution above a critical value of
$\rho^{(3D)} \lambda^3$, which is the signature for BEC. The
absence of BEC in two dimensions remains valid for an interacting
gas with repulsive interactions \cite{Mermin66,Hohenberg67}.

In spite of the absence of BEC, the physics of an interacting,
homogeneous 2D gas is still very rich. One can show that at
sufficiently low temperature the gas has a superfluid component of
density $\rho_s$ \cite{Berezinskii,KT73,Nelson77}. When
temperature increases, both the thermal wavelength $\lambda$ and
the superfluid fraction decrease until the product $\rho_s
\lambda^2$ reaches the critical value
 \begin{equation}
 T = T_c\ :\qquad \rho_s \lambda^2=4\ .
 \label{BKT}
 \end{equation}
The superfluid density then suddenly drops to zero, and the gas
becomes normal \cite{Nelson77}. The phase transition also
manifests itself in the 1-body correlation function for the atomic
field $g_1(r)=\langle \psi^*(\vec r)\;\psi(0)\rangle$. Below $T_c$
one expects a power law decay:
 \begin{equation}
T \leq T_c\ :\qquad g_1(r) \propto r^{-1/(\rho_s \lambda^2)}\ ,
 \label{powerlawdecay}
 \end{equation}
reaching  $g_1(r) \propto r^{-1/4}$ for $T=T_c$, whereas $g_1(r)$
decays exponentially for $T>T_c$.

We now discuss the physical phenomenon at the origin of this phase
transition \cite{Berezinskii,KT73}. We use a simple modeling of
the system, which consists in neglecting density fluctuations of
the gas to deal only with phase fluctuations. At low temperature
the phase varies smoothly with position and can be expanded in
normal modes. This amounts to consider only phonon excitations of
the system. Using the equipartition theorem, one can derive the
algebraic decay of the $g_1$ function given in
(\ref{powerlawdecay}). When the temperature increases, other
excitations like quantized vortices start to play a significant
role. A quantized vortex is a point where the density is zero and
around which the phase rotates by a multiple of $2\pi$. For our
purpose it is sufficient to consider only singly charged vortices,
where the phase varies by $\pm 2 \pi$ around the vortex core. The
Berezinskii-Kosterlitz-Thouless phase transition, occurring at
$T=T_c$ (see (\ref{BKT})), corresponds to the following mechanism:
 \begin{itemize}
 \item
For $T<T_c$ the free vortices are absent. Vortices exist in the
system only in the form of bound pairs, formed by two vortices
with opposite circulations. The contribution of these vortex pairs
to the decay of the correlation function $g_1$ is negligible, and
the algebraic decay of $g_1$ is dominated by the phonons.

 \item
For $T>T_c$, the free vortices form a disordered gas of phase
defects, which are responsible for the exponential decay of $g_1$.
\end{itemize}

Neglecting density fluctuations allows for a qualitative
understanding of the BKT transition, but does not provide the
relation between the total density $\rho$ and the superfluid
density $\rho_s$. This question has been addressed in
\cite{Fisher88} in the limit of an ultra weakly interacting gas
($\ln(\ln(1/\tilde g))\gg 1$). The case of weak, but more
realistic interactions, has been addressed numerically in
\cite{Prokofev01} and gives $\rho_s/\rho=4/\ln(\zeta/\tilde g)
\sim 0.5$ for our parameters (the dimensionless number $\zeta$ is
found numerically  $\simeq 380$).

\section{The trapped gas}
 \label{sec:trapped}
Consider now a gas of $N$ particles confined by the potential
$V(\vec r)=M(\omega_x^2 x^2+\omega_y y^2)/2$ in the $xy$ plane.
The presence of a trap modifies the density of states and BEC is
predicted to occur for an ideal gas when the temperature is below
the critical temperature $T_0$ \cite{Bagnato91}:
 \begin{equation}
N=\frac{\pi^2}{6}\; \left( \frac{k_BT_0}{\hbar \bar\omega}
\right)^2 \qquad \qquad \bar \omega^2=\omega_x\omega_y\ .
 \label{threshold2D}
 \end{equation}
However it should be pointed out that condensation is a very
fragile phenomenon in a 2D harmonic potential. To illustrate this
we calculate the spatial density $\rho(\vec r)$ using the
semiclassical approximation. Starting with the expression for the
phase space density
 \[
w(\vec r,\vec p)=\frac{1}{h^2}\;
\frac{1}{\exp((\frac{p^2}{2m}+V(\vec r)-\mu)/k_BT) -1}\ ,
 \]
we obtain
\[
\rho(\vec r)=\int w(\vec r,\vec
p)\;d^2p=-\frac{1}{\lambda^2}\ln(1-e^{(\mu-V(\vec r))/k_BT})\ .
 \]
Taking $\mu \to 0$ to reach the condensation threshold, we find
 \[
\rho_{\rm max}(\vec r)\;\lambda^2=-\ln(1-e^{- V(\vec r)/k_BT_0})\
.
 \]
Integrating $\rho_{\rm max}(\vec r)$ over the whole space, we
recover the threshold (\ref{threshold2D}). However, we also note
that $\rho_{\rm max}(0)=\infty$. This suggests that condensation
in a 2D harmonic potential requires an infinite 2D spatial density
at the trap center, a criterion which cannot be fulfilled as soon
as repulsive interactions play a significant role. Note that this
is in sharp contrast with the case of a 3D harmonically trapped
Bose gas, where condensation occurs when $\rho_{\rm
max}^{(3D)}(0)\simeq 2.612/\lambda^3$, which is finite and (within
the semi-classical approximation) equal to the threshold density
in a homogenous system. We refer the reader to
\cite{Fernandez02,Holzmann05} for a detailed discussion of this
subject and a collection of recent references.

In the limit of large atom numbers and low temperature, we can
still determine the equilibrium shape of the gas by looking at the
balance between the trapping potential and the repulsive
interatomic potential (Thomas-Fermi approximation). We assume that
the atomic gas forms a quasi-condensate
\cite{Popov72,Kagan87,Prokofev01,Andersen02}, so that $\langle
(\hat \psi^\dagger(\vec r))^2 (\hat \psi(\vec r))^2 \rangle \simeq
\left(\langle \hat \psi^\dagger(\vec r) \hat \psi(\vec r)
\rangle\right)^2= \rho_c^2(\vec r)$. We then get an atomic
distribution varying as an inverted parabola
 \[
\rho_c(\vec r)= \rho_c(0) \;
\left(1-\frac{x^2}{R_x^2}-\frac{y^2}{R_y^2} \right)\qquad \qquad
\frac{\hbar^2}{M}\;\tilde g\;\rho_c(0)=\mu\ ,
 \]
where the chemical potential $\mu$ and the radii of the clouds
$R_i$, $i=x,y$ are:
 \[
 \mu=\hbar \bar \omega \; ( N\tilde g/\pi)^{1/2}
 \qquad \qquad
\omega_iR_i=\sqrt{2\mu/M}\ .
 \]

In the regime where $N\tilde g \gg 1$, the chemical potential is
much larger than $\hbar \bar \omega$ and the dynamics of the gas
at low energy is dominated by phonons. One then recovers several
features of the homogeneous 2D gas \cite{Petrov04}. In particular
for $\xi\ll r\ll R_x,R_y$, the correlation function
$g_1(r)=\langle \psi^\dagger (\vec r)\;\psi(0)\rangle$ decays
algebraically at low temperature with exponent
$1/(\rho_c(0)\lambda^2)$ (see eq. (2.59) in \cite{Petrov04}). A
crossover to a true BEC occurs when the decay of $g_1$ over the
sizes $R_x,R_y$ of the quasi-condensate is negligible.

Harmonically trapped gaseous samples are finite in size and have
non-uniform density. For these reasons, one does not expect to
observe a sharp BKT transition to superfluidity, but rather a
smoothed out crossover. However, an interesting question is
whether, at least in principle, in the thermodynamic limit one
could observe a signature of a sharp transition in a harmonic
trap. For simplicity we consider an isotropic trapping
($\omega_x=\omega_y=\bar \omega$) and define the thermodynamic
limit by taking  $N \rightarrow \infty$ and $\omega \rightarrow
0$, while keeping $N \omega^2$ constant. We can then duplicate the
Kosterlitz-Thouless argument \cite{KT73} to find the probability
for having a free vortex appearing close to the center of the
quasi-condensate under the influence of thermal fluctuations.  We
choose as a sample region the central disk of radius $\epsilon R$,
with $\epsilon \ll 1$ and kept constant as $R \rightarrow \infty$.
The kinetic energy of the gas due to a free vortex in the center
of the trap is related to the superfluid density $\rho_s(\bs r)$,
which we assume to be close to the quasi-condensate density
$\rho_c(\bs r)$ \cite{rhosvsrhoc}
 \[
E_K=\int\frac{1}{2}\;\left( \frac{\hbar}{Mr} \right)^2\;
\rho_s(r)\; d^2r \simeq \frac{\pi
\hbar^2}{M^2}\;\rho_s(0)\;\ln(R/\xi)\ .
 \]
The entropy associated with the possible choices for the location
of a free vortex of size $\xi$ in the disk of radius $\epsilon R$
is $S=k_B\;\ln((\epsilon R)^2/\xi^2)= 2k_B\;\left[\ln(R/\xi) +
\ln(\epsilon)\right]$. Therefore in the limit $R/\xi \gg
1/\epsilon \gg 1$, the free energy $F=E_K-TS$ associated with a
free vortex in the central disk of radius $\epsilon R$ varies as
 \begin{equation}
\frac{F}{k_B T}= \frac{1}{2}\left(\rho_s(0)\lambda^2-4
\right)\;\ln(R/\xi)\ .
 \label{free_energy}
 \end{equation}
We thus expect a rapid variation of the average number of vortices
present in the central region when $\rho_s (0)\lambda^2$ varies
around 4. When $F>0$ (i.e. $\rho_s(0) \lambda^2 >4$) the
probability to get a free vortex in this region is small, as the
energy cost for a free vortex dominates over the gain in entropy.
On the contrary for $F<0$, i.e. $\rho_s(0) \lambda^2 <4$, we
expect a proliferation of free vortices around the center of the
trap and a jump of the local superfluid density to zero, in
analogy with the homogenous system. Finally, note that even for
$\rho_s(0) \lambda^2 >4$, free vortices can appear in the
periphery of the condensate, because the energy cost for a
off-centered vortex is lower. The role of these outer vortices on
the coherence properties at the center of the condensate remains
to be studied.

\section{Experimental study}
 \label{experiment}

\subsection{Matter wave heterodyning}

In order to study the phase coherence of our quasi-2D gases, we
have implemented a matter-wave heterodyning technique. When all
confining potentials are suddenly turned off the two clouds expand
predominantly perpendicular to the $xy$ plane. As they overlap, a
3D matter wave interference pattern forms with a spatial period
$D=2\pi \hbar t/(md)$ along the $z$ direction, where $t$ is the
time-of-flight (TOF) duration \cite{Andrews97}. After $t=20$~ms,
the projection of the 3D interference pattern onto the $xz$ plane
is recorded on a camera, using a resonant probe laser directed
along $y$.

\begin{figure}
 \centerline{\includegraphics{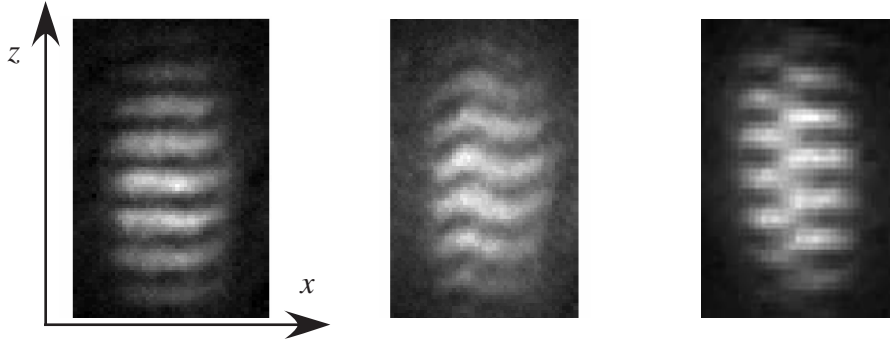}}
 \caption{Examples of interference patterns between two quasi 2D Bose gases.
 Left: low temperature pattern; Middle: high temperature pattern; Right: example
 of a dislocated pattern revealing the presence of a vortex in one of the two planes.}
 \label{fig:interf}
\end{figure}

Three examples of interference patterns are shown in
fig.~\ref{fig:interf}. The left pattern is typical of what is
obtained at very low temperature, with nearly straight fringes.
The middle pattern has been obtained for a larger temperature and
the presence of phase fluctuations in the two planes is revealed
by the waviness of the interference fringes. Finally we
occasionally observe patterns with one (or several)
dislocation(s), such as the one shown in the right of
fig.~\ref{fig:interf}. We interpret such a dislocation as the
signature of a free vortex in one of the two atomic clouds
\cite{Inouye01,Chevy01}. The matter wave heterodyning technique
thus offers a way to study both long wavelength excitations
(phonons) by analyzing the smooth variations of the fringes of the
interference patterns, and the point-like defects such as free
vortices by looking for sharp dislocations in the fringes.

\subsection{Extracting $g_1$ from interference patterns}
We use in this paragraph a simple modeling, where we assume that
the two gases have the same uniform amplitude $\psi_0$ and
independently fluctuating phases $\varphi_a (x,y)$ and
$\varphi_b(x,y)$. The interference signal recorded after TOF is
  \begin{equation}
S(x,z)\propto  2\psi_0^2+e^{2i\pi z/D}\;c(x)+e^{-2i\pi
z/D}\;c^*(x)  \label{modeling_interf}
 \end{equation}
with
 \begin{equation}
 c(x)=\frac{\psi_0^2}{L_y}\;\int_{-L_y/2}^{L_y/2}
 e^{i (\varphi_a(x,y)-\varphi_b(x,y)}\;dy\ .
   \end{equation}
At this stage the integration distance $L_y$ is arbitrary, and it
can differ from the total length $2\,R_y$ along the $y$ direction.
For a quantitative analysis of long-range correlations as a
function of temperature, we adopt the method proposed in
\cite{Polkovnikov06}. The idea is to integrate the coefficient
$c(x)$ appearing in (\ref{modeling_interf}) over a variable length
$L_x$:
 \[
C(L_x)=\frac{1}{L_x}\int_{-L_x/2}^{L_x/2} c(x)\; dx
 \]
and average $|C(L_x)|^2$ over many images recorded in the same
conditions. Using the fact that the phases $\varphi_a$ and
$\varphi_b$ are uncorrelated, we obtain:
 \begin{eqnarray}
\langle |C(L_x)|^2 \rangle &=&\frac{1}{L_x^2}\int \hskip -2mm\int
\langle c(x)\;c^*(x') \rangle\;dx\;dx'\nonumber\\
&=& \frac{\psi_0^2}{L_x^2 L_y^2} \int \hskip-2mm\int
\hskip-2mm\int\hskip-2mm\int \langle
e^{i(\varphi_a(x,y)-\varphi_a(x',y')}\rangle\;
 \langle
 e^{i(\varphi_b(x',y')-\varphi_b(x,y)}\rangle\;dx\,dx'\,dy\,dy' \nonumber\\
 &\simeq& \frac{1}{L_x} \int_{-L_x/2}^{L_x/2} |g_1(x,0)|^2\;dx
  \quad \propto\ \left( \frac{1}{L_x} \right)^{2\alpha} \ ,\label{decay_C}
 \end{eqnarray}
where we have assumed $L_x\gg L_y$. The long-range physics is then
captured in a single parameter, the exponent $\alpha$. It is
straightforward to understand the expected values of $\alpha$ in
some simple cases. In a system with true long-range order, $g_1$
would be constant and the interference fringes would be perfectly
straight. In this case  $\alpha = 0$, corresponding to no decay of
the contrast upon integration. In the low temperature regime,
where $g_1$ decays algebraically (see (\ref{powerlawdecay})) the
exponent $\alpha$ coincides with that of $g_1$. In the high
temperature case, where $g_1$ decays exponentially on a length
scale much shorter than $L_x$, the integral in (\ref{decay_C}) is
independent of $L_x$. In this case $\alpha = 0.5$, corresponding
to adding up local interference fringes with random phases. The
BKT mechanism corresponds to a transition between a power law with
exponent ($1/(\rho_s\lambda^2)\leq 0.25$) to an exponential decay
of $g_1$. It should thus manifest itself in a uniform system as a
sudden jump of $\alpha$ from 0.25 to 0.5 when the temperature
varies around $T_c$.

\begin{figure}
 \centerline{\includegraphics[width=14cm]{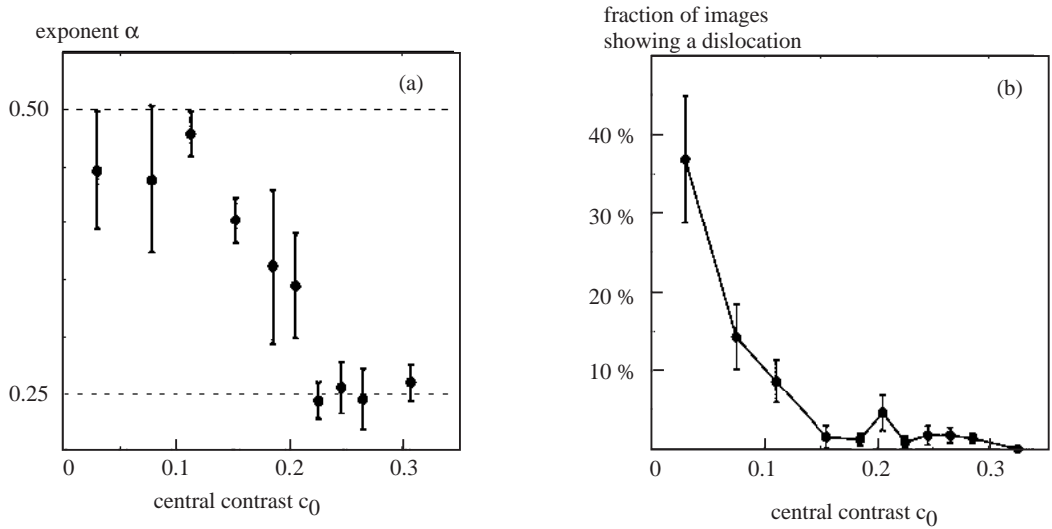}}
 \caption{(a) Fitting exponent $\alpha$ as a function of the
 average central contrast $c_0$. (b) Fraction
 of images showing at least one dislocation as a function of $c_0$.}
 \label{fig:results}
\end{figure}

\subsection{Experimental results}
In order to apply the method of \cite{Polkovnikov06} to our data,
we fit the density distribution recorded after TOF with a function
 \[
F(x,z)=G(x,z)\left[ 1+c(x)\;\cos\left( 2\pi z/D + \theta(x)
\right) \right]\ ,
 \]
where $G(x,z)$ is a gaussian envelope, and we proceed with the
experimental $c(x)$ as discussed above. We average $ |C(L_x)|^2$
over $\sim 100$ images and fit it with the function
$L_x^{-2\alpha}$. We have plotted in fig.~\ref{fig:results}a the
results for $\alpha$, as a function of the average central
contrast $c_0$ of the images. We choose the latter quantity as a
temperature label since we feel that it is less model-dependent
than other estimates, based for example on the tails of the atomic
TOF distribution. Starting at high temperatures, for values of
$c_0$ ranging from a few \% (experimental limit of detectability)
up to about 13\%, $\alpha$ is approximately constant and close to
0.5. When the temperature is reduced further, $\alpha$ rapidly
drops to about 0.25, and for even lower temperatures (larger
$c_0$) it levels off. We thus clearly observe a transition between
two qualitatively different regimes at high and low temperatures.
The values of $\alpha$ above and below the transition are in
agreement with the theoretically expected jump at the BKT
transition in a uniform system.

Figure~\ref{fig:results}b shows the number of sharp dislocations
at different temperatures. For the count we consider only the
central, 30~$\mu$m wide region of each image, which is smaller
than the length of our smallest quasi-condensates. We observe a
clear onset of dislocation proliferation with increasing
temperature, and this onset coincides with the loss of quasi-long
range coherence described by the rapid variation of $\alpha$.

We can also independently estimate the cloud's temperature and
quasi-condensate density at the onset of quasi-long-range
coherence. For images with $c_0=0.15$, we get $\rho_c\lambda^2 = 6
\pm 2$ \cite{Hadzibabic06}. This is in agreement with the BKT
theory, and we refer the reader to \cite{Prokofev01} for a further
discussion on the difference between $\rho_c$ and $\rho_s$.

\subsection{Discussion and summary}

The experimental procedure presented here constitutes a powerful
technique to study the properties of quasi-2D atomic gases at
ultra-low temperature. Our results show that, in agreement with
the prediction of the BKT theory, there exists a relatively narrow
temperature domain over which the coherence of our trapped planar
Bose gas varies rapidly, with the exponent $\alpha$ in
(\ref{decay_C}) switching from $\approx 0.25$ (low temperature
side) to $\approx 0.5$ (high temperature side). In addition the
matter wave heterodyning technique allows for the direct
observation of dislocations of the interference pattern due to
vortices, and it supports the notion that the unbinding of
vortex-antivortex pairs is the microscopic mechanism destroying
the quasi-long range coherence in 2D systems.

However, we should point out that the quantitative agreement
between our results and the predictions of \cite{Polkovnikov06}
derived from BKT theory might be partly fortuitous. Indeed though
we concentrated on the quasi-uniform part of the images, the
geometry effects in our elongated samples ($R_x\sim 12 R_y$) could
still be important. In particular the integration length $L_y$ in
our experiment is equal to the full width of the system $2R_y$ in
that direction. The condition $L_x \gg L_y$ used to derive
(\ref{decay_C}) then implies that we probe the planar system on a
distance scale $L_x$ that is large compared to its other dimension
$2R_y$. Over this length scale the system acquires some quasi-one
dimensional features and $g_1$ is expected to decay exponentially,
possibly with a very large coherence length (see e.g.
\cite{Gerbier04}). If this is indeed the case, our experiment is
detecting a rapid change of the decay length of the exponential,
rather than a change in the functional form of $g_1$ given in
(\ref{powerlawdecay}). The elongated nature of our samples also
changes the quantitative calculation of the threshold for the
apparition of a vortex (see section~\ref{sec:trapped}). Finally we
note that we detect here only a subset of vortices, i.e. those
which are well isolated and close to the center of the cloud. The
heterodyning technique is not well adapted to the detection of
large numbers of vortices, that are expected for temperatures
notably above $T_c$. In the latter case indeed, the fringe pattern
becomes hardly detectable and thermally activated phonon modes
with a very short wavelength along $x$ can also have a significant
contribution. Our work should then be considered as a first step
in the investigation of the BKT mechanism in planar atomic gases.
A quantitative theoretical analysis of our system will probably
require a numerical study, for example along the lines of
\cite{Simula06}. In parallel, future experimental studies could be
performed with more isotropic samples, and could use a ``slicing
method" similar to \cite{Andrews97} in order to extract local
information from the center of the cloud.

\begin{theacknowledgments}
We thank E. Altman, E. Demler, M. Lukin, A. Polkovnikov, P.-S.
Rath, D. Stamper-Kurn, and S. Stock for useful discussions. We
acknowledge financial support by IFRAF, ACI nanonscience, ANR, the
Alexander von Humboldt foundation (P.K.) and the EU under
Marie-Curie Fellowships (Z.H. and P.K.). Laboratoire Kastler
Brossel is a research unit of Ecole normale supérieure and
Université Paris 6, associated to CNRS.
\end{theacknowledgments}


\begin{thebibliography}{99}
\bibitem{rhosvsrhoc}
The distinction between superfluid density $\rho_s$ and
quasi-condensate density $\rho_c$ is delicate. We refer the reader
to \cite{Prokofev01}, where the difference between the two
quantities is discussed.
\end{thebibliography}
\end{document}